# Morphological and electrical properties of Nickel based Ohmic contacts formed by laser annealing process on n-type 4H-SiC

S. Rascunà[1*], P. Badalà[1], C. Tringali[1], C. Bongiorno[2], E. Smecca[2], A. Alberti[2], S. Di Franco[2], F. Giannazzo[2], G. Greco[2], F. Roccaforte[2], M. Saggio[1]

[1] *STMicroelectronics SRL, Stradale Primosole 50, 95121 Catania, Italy*

[2] *Consiglio Nazionale delle Ricerche – Istituto per la Microelettronica e Microsistemi (CNR-IMM), Strada VIII, n.5 Zona Industriale, I-95121 Catania, Italy*

(*) Corresponding author: simone.rascuna@st.com,

**Abstract.** This work reports on the morphological and electrical properties of Ni-based back-side Ohmic contacts formed by laser annealing process for SiC power diodes. Nickel films, 100 nm thick, have been sputtered on the back-side of heavily doped 110 µm 4H-SiC thinned substrates after mechanical grinding. Then, to achieve Ohmic behavior, the metal films have been irradiated with an UV excimer laser with a wavelength of 310 nm, an energy density of 4.7 J/cm$^2$ and pulse duration of 160 ns. The morphological and structural properties of the samples were analyzed by means of different techniques. Nanoscale electrical analyses by conductive Atomic Force Microscopy (C-AFM) allowed correlating the morphology of the annealed metal films with their local electrical properties. Ohmic behavior of the contacts fabricated by laser annealing have been investigated and compared with the standard Rapid Thermal Annealing (RTA) process. Finally, it was integrated in the fabrication of 650V SiC Schottky diodes.

## Introduction

Silicon carbide (4H-SiC) is one of the key materials to fabricate high-power and low ON-resistance ($R_{ON}$) devices in the next generation of power electronics systems [1, 2]. One of the most important issues in SiC technology is the fabrication of reliable low-resistance back-side Ohmic contacts to SiC devices.

For the n-type SiC, annealed Ni-films are commonly used to form nickel silicide ($Ni_2Si$) back-side Ohmic contacts. Typically, rapid thermal annealing (RTA) exceeding 900°C are used to achieve an Ohmic behavior [3]. However, today there is the need to replace the conventional thermal annealing by laser annealing processes carried out on the back-side of thinned wafers at the end of the fabrication flow [4].

In Fig. 1, a schematic of a standard flow chart (with and without grinding step) for the fabrication of Junction Barrier Schottky (JBS) diode is shown in comparison with a new laser annealing process flow. In particular, in the standard manufacturing process flow (Fig. 1a,b), a high temperature RTA process (900 - 1000 °C) is typically required for the formation of the $Ni_2Si$ back-side Ohmic contact [5]. This process must be carried out before defining the front side metal to prevent undesired interface reactions and electrical degradation of the Schottky barrier. Hence, this standard silicidation process represents today a technological bottleneck, due to the complexity of the process flow integration that requires multiple flip over of the wafers and reliability of back-side layer due to the exposition to many processes remaining in the flow.

In this context, the introduction of wafer grinding, to reduce the resistive contribution of the substrate (Fig.1b), has to deal with consequent limitation in step processes and the increasing risk of wafer breakage.

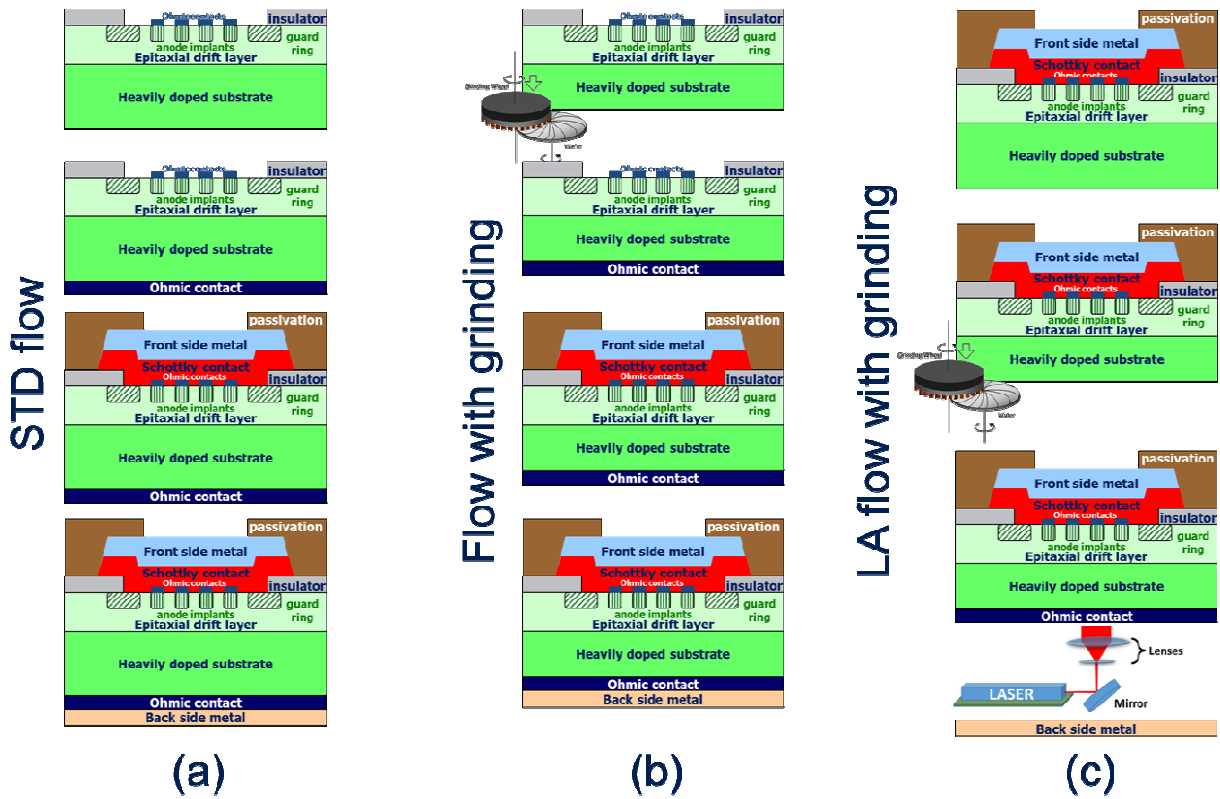

Fig. 1. Schematic flow charts for the fabrication of a JBS diode: standard wafers with back-side ohmic contact by RTA process (a), thinned wafers with back-side ohmic contact by RTA process (b), thinned wafers with back-side ohmic contact by laser annealing process (c).

However, such process can bring great advantages in terms of device power dissipation. In fact, as can be seen in the calculation shown in Fig. 2a, in a 650 V SiC Schottky diode fabricated onto a 350 µm thick substrate, about 70 % of the total $R_{ON}$ is represented by the SiC substrate contribution, $R_{sub}$ [6]. On the other hand, thinning the substrate to 110 µm allows reducing this resistive contribution down to 44 % of the total $R_{ON}$ (Fig. 2b) [6]. This latter explains why for medium voltage applications (600-1200 V), the wafer grinding step has become mandatory in SiC technology to reduce the substrate thickness and, hence, to minimize the total device $R_{ON}$.

From this point of view, laser annealing (Fig. 1c) represents an alternative valid solution for achieving the silicidation with a limited heat transfer. Moreover, the use of laser annealing in back Ohmic contact formation enables the possibility to complete the device front side first, and then to process the back-side contact without detrimental effects for the Schottky barrier and without limitation on the thinning of the wafers. The possibility to obtain Ohmic contact formation by the use of laser annealing has been already demonstrated in Silicon [7, 8, 9]. In the recent years, several works reported on the formation of Ohmic contacts to SiC thinned substrate using laser annealing (LA) processes, suitable both for diode and MOSFET technology [4, 10, 11, 12, 13, 14, 15].

In this paper a morphological, structural and electrical characterization of Ni-based back-side Ohmic contacts formed by laser annealing on 4H-SiC has been reported. In particular, the electrical behavior showed the possibility to obtain an Ohmic behavior under certain process condition. In these cases, formation of nickel silicide has been confirmed by Transmission Electron

Microscopy (TEM) analysis and X-ray diffraction (XRD) analysis. Moreover, a combination of nanoscale morphological and electrical analysis allowed correlating the occurred Ni-SiC reaction with the achieved Ohmic contact behavior.

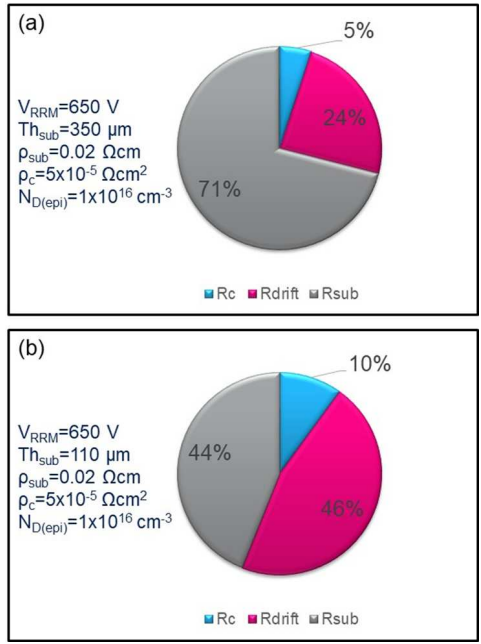

Fig. 2. Resistive contributions (in terms of contact resistance, $R_c$, epitaxial-layer resistance, $R_{drift}$, and substrate resistance, $R_{sub}$) of the total $R_{ON}$ in a 650 V 4H-SiC Schottky diode for two different substrate thickness of 350 µm (a) and 110 µm (b). For this calculation, substrate resistivity of 0.02 Ωcm, specific contact resistance of backside contact of $5\times10^{-5}$ Ωcm$^2$, epilayer doping of $1\times10^{16}$ cm$^{-3}$ have been used.

**Experimental**

Nickel films, 100 nm thick, have been sputtered on the back-side of heavily doped 4H-SiC substrates, whose thickness had been reduced down to 110 µm by mechanical grinding. Then, the metal films have been irradiated with an ultraviolet (UV) excimer laser with a wavelength of 310 nm, an energy density of 4.7 J/cm$^2$ and pulse duration of 160 ns. Patterned samples, as Transmission Line Model (TLM) structure, have been defined to evaluate the electrical properties of the contacts. The current voltage (I-V) measurements on the TLM test patterns were acquired using a semiconductor device parameter analyzer (Agilent B1500A). The back-side morphologies and microstructures have been investigated by Scanning Electron Microscopy (SEM – Fei 865 dual beam) and Transmission Electron Microscopy (TEM – Jeol JEM 2010F), while X-ray diffraction patterns were acquired using a D8 Discover (Bruker AXS) diffractometer equipped with a Cu source and thin films attachment to obtain information about the crystalline structure of materials. Nanoscale electrical analyses by conductive Atomic Force Microscopy (C-AFM – DI3100 with Nanoscope V controller) using Pt coated Si tips allowed to correlate the morphology of the annealed metal films with their local electrical properties. Finally, the laser annealing process has been integrated in the fabrication of a 4H-SiC power JBS diode, and the I-V characteristics of this diode were collected using a high power curve tracer (Sony Tektronix 371A).

**Results and discussion**

Electrical measurements on dedicated test pattern have been performed on the heavily doped substrate surface (0.02 Ωcm). As can be observed in Fig. 3, linear I-V characteristics measured on adjacent pads indicated the achievement of the Ohmic behavior. On the other side, Ni silicide, formed by standard RTA annealing (1000 °C), exhibits a higher current, indicating a better contact resistance. It must be pointed out that the test patterns used for the electrical measurements have been fabricated on the back-side of thick (350 µm) heavily doped substrates. Hence, the lack of a vertical isolation of the TLM patterns, together with the low sheet resistance of the substrate, did not allow to quantitatively determining the value of the specific contact resistance.

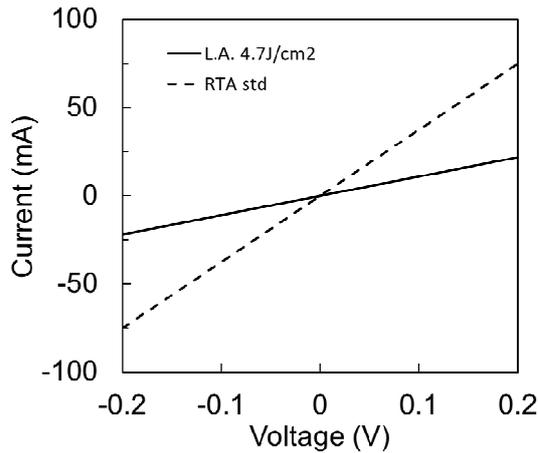

Fig. 3. I-V curves acquired on adjacent TLM patterns of Ni-based Ohmic contacts formed by standard RTA (1000°C) and Laser annealing processes (at an energy density of 4.7J/cm$^2$).

A deep structural analysis of the LA treated Ni films have been performed. In particular, X-ray diffraction analyses (Fig. 4) revealed the formation of the Ni$_2$Si phase [16, 17] with main peaks as those detected in the reference sample (RTA). The comparison of acquisitions by symmetric diffraction (2theta-omega) and by grazing incidence highlights a wide distribution of the growth axes in the Ni$_2$Si polycrystals array. As a further insight, the large full width at half maximum of the peaks compared to the reference case, that causes the peak to overlap, marks the presence of a fine-grained material as reaction product.

Moreover, after laser treatment, two different regions can be identified. In fact, Ni agglomeration has been observed by SEM analysis, as clearly visible in Fig. 5. Here, the Ni agglomerated in characteristic islands, which leave uncovered other semiconductor regions.

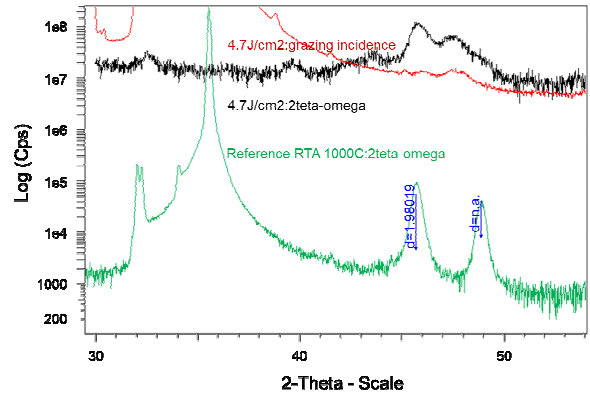

Fig. 4. XRD pattern of the back-side layer after laser treatment (black line) compared with a reference sample (RTA): (green line) grazing incidence profile; (red line) 2theta-omega profile. The labelled peak positions in blue identify the orthorhombic Ni$_2$Si phase.

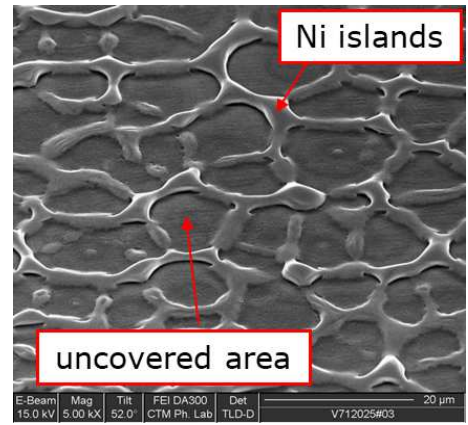

Fig. 5. SEM in plan view of the back-side layer after 100 nm Ni sputtering and 310 nm laser treatment at an energy density of 4.7 J/cm$^2$.

Both Ni islands (2) and areas uncovered by nickel (1) have been further investigated by TEM analysis (Fig. 6a). The Ni islands observed in figure 5 are composed of nickel silicide, in a Ni$_2$Si crystal phases, mixed with random distributed spherical carbon clusters. Moreover, a continuous layer of C cluster has been also detected at the interface with SiC, as typically in standard RTA Ni-based contacts. On the other side, structure and chemical distribution of the uncovered areas result more complex and unexpected. From TEM image of Fig. 6b we observe a 350nm thick defective region, composed of three different layers, a 10 nm continuous layer on the surface (1), an intricate network of crystal grains and spherical clusters in the middle (2) and a crystalline SiC layer with a high density of fault in the stacking sequence (3).

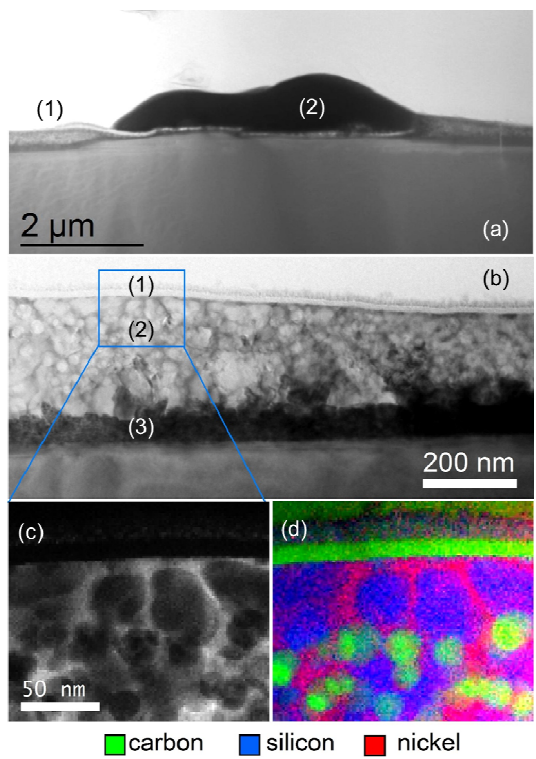

Fig. 6. TEM images of the laser treated Ni/SiC interface, which leads to uncovered area (a-1) and Ni agglomeration (a-2). Three different layers in the uncovered regions are identified (b-1, b-2, b-3). HAADF STEM image (c) together with map of elements (d), of the marked area in fig b.

The composition of the first two layers in the uncovered area were revealed by using scanning TEM configuration and electron energy loss (EEL) spectrum image acquisition. Thanks to the parallel acquisition of the high angle annular dark field image (HAADF) and EEL spectrum, point by point, two dimension chemical maps of the elements present in the marked area in Fig 6b has been obtained. Three different elemental maps (carbon, silicon and nickel) have been extracted and over imposed in the RGB image showed in Fig. 6d. In this way it is possible to recognize carbon accumulation on the surface (green continuous layer), pure silicon grains (blu) and pure carbon sphere (green) enveloped by a network containing nickel (violet).

Fig. 7a shows a schematic of the C-AFM experimental setup employed for high resolution mapping of current injection through the LA Ni back-contact. A representative morphology and the corresponding current map acquired on a large scan area (50 µm × 50 µm) are reported in Fig. 7b and Fig. 7c, respectively. The same characteristic pattern found in the SEM image of Fig. 5, composed by regions covered and uncovered with Ni, can be observed in the C-AFM topography. From the comparison of the morphological and electrical maps, it is very interesting to note that the injected current levels in the uncovered regions are similar to those measured on the Ni covered islands. To better visualize the uniformity of current injection in the Ni uncovered areas, higher resolution morphology and current maps in these regions are reported in Fig. 7d and Fig. 7e. This observation clarifies that the entire back-contact area contributes to the total current density flowing through the device, independent of the elemental distribution. Furthermore, it indicates that the Ohmic behavior of the LA back contact is guaranteed by the thermal reaction needed to form the Nickel Silicide rather than the $Ni_2Si$ itself.

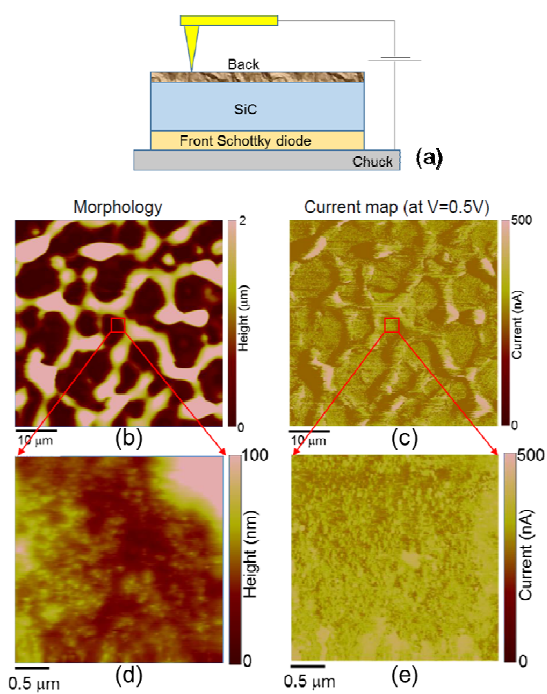

Fig. 7. Schematic of C-AFM probe (a). Morphology map (b, d) and current map(c, e).

Finally, the investigated contacts have been integrated in the fabrication of 650V JBS SiC diodes. The laser annealing process gives the possibility to handle a 110 µm thinned SiC wafer. Fig. 8 shows the I-V characteristic at 175 °C of the SiC JBS diodes processed with laser annealing (110 µm thick) in comparison with those obtained by standard RTA process (180 µm thick). Despite of the better electrical behavior observed in the case of contacts processed by standard RTA treatments (see Fig. 3) the benefit on the reduction of the substrate resistive contribution is clearly visible. Then, the gain in the forward voltage drop $V_F$ at high current level represents indirectly the benefit in terms of surge current ruggedness of diodes on thinned substrate.

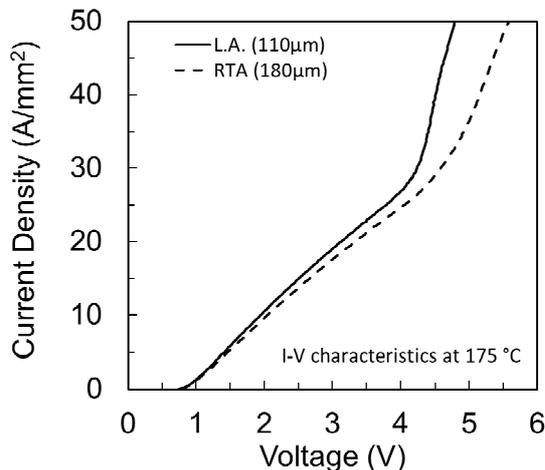

Fig. 8. Forward I-V characteristic at 175 °C of 650 V SiC JBS diodes with different substrate thickness (110 µm vs 180 µm) and back-side contact process (Laser annealing vs RTA).

**Conclusion**

This paper reports a morphological, structural and electrical analysis of Ni-based Ohmic contacts formed by a laser annealing process on 4H-SiC.

The back-side Ohmic contact on heavily doped 4H-SiC substrates has been formed by a laser process with energy density of 4.7 J/cm$^2$ and pulse duration of 160ns. Ni based agglomeration and areas uncovered by nickel on the bask-side surface have been observed and analyzed after the laser treatment as an indication that SiC melting point has been locally reached. Interestingly, C-AFM demonstrated that both areas contribute to the total current density in the device. Finally, the use of Laser Annealing technology is justified by the possibility to use grinding process down to 110 µm. In fact, the advantage to use grinding substrate, with a reduced resistive contribution, processed by Laser Annealing has been proved on a 650V JBS diode. However, additional processing optimizations are still required to achieve Ohmic contacts with electrical properties comparable to the standard RTA process.

**Acknowledgements**

This work was carried out in the framework of the ECSEL JU project WInSiC4AP (Wide Band Gap Innovative SiC for Advanced Power), grant agreement n. 737483.

**References**

[1] T. Kimoto, J.A. Cooper, *Fundamentals of Silicon Carbide Technology: Growth, Characterization, Devices and Applications*, JohnWiley & Sons Singapore Pte. Ltd., Singapore, 2014.
[2] F. Roccaforte, P. Fiorenza, G. Greco, R. Lo Nigro, F. Giannazzo, F. Iucolano, M. Saggio, *Emerging trends in wide band gap semiconductors (SiC and GaN) technology for power devices*, Microelectronic Engineering, 187-188, 66 (2018).
[3] F. Roccaforte, F. La Via, V. Raineri, *Ohmic contacts to SiC*, Int. J. High Speed Electronics and Systems 15, (2005) 781-820.
[4] R. Rupp, R. Kern, R. Gerlach, Proc. of ISPSD2013, pagg. 51-54, (2013).
[5] F. Roccaforte, M. Vivona, G. Greco, R. Lo Nigro, F. Giannazzo, S. Rascunà, M. Saggio, *Metal/Semiconductor Contacts to Silicon Carbide: Physics and Technology*, Mater. Sci. Forum vol. 924, (2018) pp. 339-344.
[6] F. Roccaforte, G. Brezeanu, P.M. Gammon, F. Giannazzo, S. Rascunà, M. Saggio, *Schottky contacts to Silicon Carbide: Physics, Technology and Applications*, in "Advancing Silicon Carbide Electronics Technology vol. I", Materials Research Foundation 37 (2018) pp.127-190 (http://dx.doi.org/10.21741/9781945291845).

[7] A. Alberti, A. La Magna, M. Cuscunà, G. Fortunato, V. Privitera, *Simultaneous nickel silicidation and silicon crystallization induced by excimer laser annealing on plastic substrate*, App. Phys. Lett. 96, 142113 (2010).

[8] A. Alberti, A. La Magna, M. Cuscunà, G. Fortunato, C. Spinella, V. Privitera, *Nickel-affected silicon crystallization and silicidation on polyimide by multipulse excimer laser annealing*, J. Appl. Phys. 108, 123511 (2010).

[9] A. Alberti, A. La Magna, *Role of the early stages of Ni-Si interaction on the structural properties of the reaction products*, J. Appl. Phys. 114, 121301 (2013).

[10] K. Nakashima, O. Eryu, S. Ukai, K. Yoshida, M. Watanabe, *Improved Ohmic Contacts to 6H-SiC by Pulsed Laser Processing*, Materials Science Forum, Vols. 338-342 (2000), pp. 1005-1008

[11] Y. Ota, Y. Ikeda, M. Kitabatake, *Laser Alloying for Ohmic Contacs on SiC at Roop Temperature*, Materials Science Forum,Vols. 264-268 (1998), pp. 783-786

[12] S. Ferrero, Albonico A., U. Meotto, G. Rombola, S. Porro, F. Giorgis, D. Perrone, L. Scaltrito, E. Bontempi, L. Depero, G. Richeiri and L. Merlin, *Phase Formation at Rapid Thermal Annealing of Nickel Contacts on C-face n-type 4H-SiC*, Materials Science Forum,Vols. 483-485 (2005), pp. 733-736

[13] A. Hürner, T. Schlegl, B. Adelmann, H. Mitlehner, R. Hellmann, A.J. Bauer, L. Frey, *Alloying of Ohmic Contacts to n-Type 4H-SiC via Laser Irradiation*, Mater. Sci. Forum 740-742, 773 (2013).

[14] T. Tabata, S. Halty, I. Toqué-Trèsonne, F. Mazzamuto, K. Huet, Y. Mori, *UV Excimer Laser Annealing for Next Generation Power Electronics*, Proc. 21st Int. Conf. Ion Implantation Technology (IIT2016) (2016) doi:10.1109/IIT.2016.7882917.

[15] F. Mazzamuto, S. Halty, H. Tanimua, Y. Mori, *Low thermal budget ohmic contact formation by laser anneal*, Mater. Sci. Forum 858, 565 (2016).

[16] A Alberti, C Bongiorno, F La Via, C Spinella, *High-resolution investigation of atomic interdiffusion during Co/Ni/Si phase transition*, Journal of applied physics 94 (1), 231-237 (2003).

[17] A Alberti, C Bongiorno, E Rimini, MG Grimaldi, *Critical nickel thickness to form silicide transrotational structures on [001] silicon*, Applied physics letters 89 (10), 102105 (2006).